\documentclass[twocolumn,english,prx,longbibliography,aps]{revtex4-1}
\pdfoutput=1
\usepackage[utf8]{inputenc}
\usepackage{amsmath}
\usepackage{amssymb}
\usepackage{graphicx}
\usepackage{booktabs}
\usepackage{url}
\usepackage{array}
\usepackage{dcolumn}



\begin{document}

\title{Compression of flow can reveal overlapping modular organization in
networks}

\author{Alcides \surname{Viamontes Esquivel}}
\email{a.viamontes.esquivel@physics.umu.se}
\affiliation{Integrated Science Lab, Umeå University, Sweden.}

\author{Martin Rosvall}
\affiliation{Integrated Science Lab, Umeå University, Sweden.}

\date{04/28/2011}
\begin{abstract}
To better understand the overlapping modular organization of large networks
with respect to flow, here we introduce the map equation for overlapping
modules. In this information-theoretic framework, we use the correspondence
between compression and regularity detection.  The generalized map equation
measures how well we can compress a description of flow in the network when we
partition it into modules with possible overlaps. When we minimize the
generalized map equation over overlapping network partitions, we detect modules
that capture flow and determine which nodes at the boundaries between modules
should be classified in multiple modules and to what degree. With a novel
greedy search algorithm, we find that some networks, for example, the neural
network of \emph{C. Elegans}, are best described by modules dominated by hard
boundaries, but that others, for example, the sparse European road network,
have a highly overlapping modular organization. 
\end{abstract}


\maketitle

\section{Introduction}

To discern higher levels of organization in large social and biological
networks
\cite{albert2002,pastor2004evolution,boccaletti2006complex,ratti2010redrawing,alba1978elite},
researchers have used hard clustering algorithms to aggregate highly
interconnected nodes into non-overlapping modules
\cite{fortunato,girvan,newman-fast} because they have assumed that each node
only plays a single modular role in a network. Recently, because researchers
have realized that nodes can play many roles in a network, they have detected
overlapping modules in networks using three approaches: a hard clustering
algorithm that is run multiple times \cite{wilkinson2004method,gfeller}; a
local clustering method that generates independent and intersecting modules
\cite{palla2007,gregory2008fast,Lancichinetti2011,whitney2011clustering}; and link clustering that
assigns boundary nodes to multiple modules
\cite{evans2009line,Ahn:2010p167,kim2011}. However, all these approaches have
limitations.The first and second approaches require several steps or tunable
parameters to infer overlapping modules and the third approach necessarily
overlaps all neighboring modules. To find simultaneously the number of modules
in a network, which nodes belong to which modules, and which nodes should
belong to multiple modules and to what degree, we use an information- theoretic
approach and the map equation\cite{rosvall2009map}.

We are interested in the dynamics on networks and what role nodes
on the boundaries between modules play with respect to flow through
the system. For example, in Fig.~\ref{fig:Illustration-of-hypothetical}(a),
Keflavik airport in Reykjavik connects Europe and North America in
the global air traffic network. When we summarize the network in modules
with long flow persistence times, should Reykjavik belong to Europe,
North America, or both? In our framework, the answer depends on the
traffic flow. That is, Reykjavik's role in the network depends on
to what degree passengers visit Iceland as tourists versus to what
degree they use Keflavik as a transit between North America and Europe.
If we assign the boundary node to both modules, for returning flow
we can increase the time the flow stays in the modules and decrease
the transition rate between the modules, but for transit flow, the
transition rate does not decrease and a single module assignment is
preferable. By generalizing the information theoretic clustering method
called the map equation \cite{RosvallBergstrom08} to overlapping
structures, we can formalize this observation and use the level of
compression of a modular description of the flow through the system
to resolve the fuzzy boundaries between modules. With this approach, modules
will overlap if they correspond to separate flow systems with shared nodes.

In the next section, we review the map equation framework, introduce
the map equation for overlapping modules, and explain how it exploits
returning flow near module boundaries. The mathematical framework
works for both generalized and empirical flow, but here we illustrate
the method by exploring the overlapping modular structure of several
real-world networks based on the probability flow of a random walker.
We also test the performance on synthetic networks and compare the
results with other clustering algorithms. Finally, in the Materials
and Methods section, we provide complete descriptions of the map equation
for overlapping modules and the novel search algorithm. 

\section{\label{sec:Results-and-discussion}Results and discussion}

\subsection{\label{sub:The-map-equation}The map equation}

The mathematics of the map equation are designed to take advantage
of regularities in the flow that connects a system’s components and
generates their interdependence. The flow can be, for example, passengers
traveling between airports, money transferred between banks, gossip
exchanged among friends, people surfing the web, or, what we use here
as a proxy for real flow, a random walker on a network guided by the
(weighted directed) links of the network. Specifically, the map equation
measures how well different partitions of a network can be used to
compress descriptions of flow on the network and utilizes the rationale
of the minimum description length principle. Quoting Peter Grünwald
\cite{grunwald2007minimum}: {}``\emph{...{[}E{]}very regularity
in the data can be used to compress the data, i.e.,~to describe it
using fewer symbols than the number of symbols needed to describe
the data literally.}" That is, the map equation gauges how successful
different network partitions are at finding regularities in the flow
on the network.

We employ two regularities for compressing flow on a network. First,
we use short code words for nodes visited often and, by necessity,
long code words for nodes visited rarely, such that the average code word
length will be as short as possible. Second, we use a two-level code
for module movements and within-module movements, such that we can
reuse short node code words between modules with long persistence times.

Because we are not interested in the actual code words, but only in
the theoretical limit of compression, we use Shannon's source coding
theorem \cite{shannon}, which establishes the Shannon entropy $H(\mathbf{p})$
as the lower limit of the average number of bits per code word necessary
to encode a message, given the probability distribution $\mathbf{p}$
of the code words,
\[
H(\mathbf{p})=-\sum_{i}p_{i}\log_{2}p_{i}.
\]

For example, if there is a message {}``ABABBAAB...'' for which the
symbols {}``A'' and {}``B'' occur randomly with the same frequency,
that is, {}``A'' and {}``B'' are independent and identically distributed,
the source coding theorem states that no binary language can describe
the message with less than $-\frac{1}{2}\log_{2}\frac{1}{2}-\frac{1}{2}\log_{2}\frac{1}{2}=1$
bit per symbol. However, if {}``A'' occurs twice as often as {}``B'',
the regularity can be exploited and the message compressed to $-\frac{1}{3}\log_{2}\frac{1}{3}-\frac{2}{3}\log_{2}\frac{2}{3}\approx0.92$
bit per symbol. To measure the per-step minimum average description
length of flow on a network, we collect the mapping from symbols {}``A''
and {}``B'' or, in our case, node names, to code words in a codebook,
and calculate the Shannon entropy based on the node-visit frequencies.

\begin{figure}[tph]
\includegraphics[width=1\columnwidth]{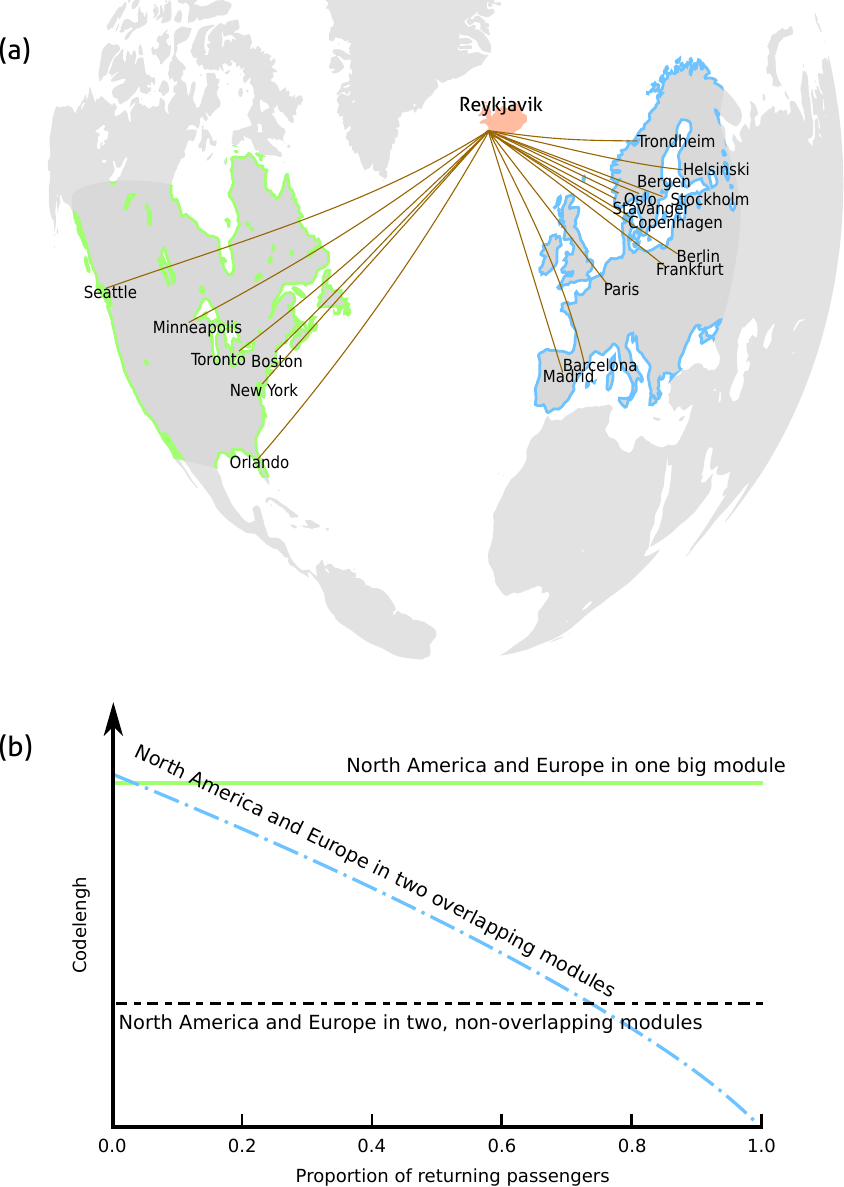}

\caption{\label{fig:Illustration-of-hypothetical}The map equation for overlapping
modules can exploit regularities in the boundary flow between modules.
The three lines in (b) show the description length as a function of
the proportion of returning passengers for three different partitions:
North America, Europe, and Reykjavik in one big module (green); North
America and Europe in two non-overlapping modules with Reykjavik in
either of the modules (black); and North America and Europe in two
overlapping modules with Reykjavik in both modules (blue).}
\end{figure}

But flow or a random walker do not visit nodes independently. For
example, if a network has a modular structure, once a random walker
enters a tightly interconnected region in the network, in the next
step she will most likely visit a node in the same tightly interconnected
region, and she tends to stay in the region for a long time. To take
advantage of this regularity and further compress the description
of the walk, we use multiple module codebooks, each with an extra
exit code that is used when the random walker exits the module, and
an index codebook that is used after the exit code to specify which
module codebook is to be used next. Now we can make use of higher-order
structure in a network. For a modular network, we can describe flow
on the network without ambiguities in fewer bits, using a two-level
code, than we could do with only one codebook, because we only use
the index codebook for movements between modules and can reuse short
code words in the smaller module codebooks. 

Given a network partition $\mathsf{M}$, it is now straightforward
to calculate the per-step minimum description length $L(\textsf{M})$
of flow on the network. We use the Shannon entropy to calculate the
average description length of each codebook and weight the average
lengths by their rates of use. For a modular partition $\textsf{M}$
with $m$ modules, the map equation takes the form:

\begin{equation}
L(\mathsf{\mathsf{M}})=q_{\curvearrowright}H(\mathcal{Q})+\sum_{i=1}^{m}p_{\circlearrowright}^{i}H\left(\mathcal{P}^{i}\right).\label{eq:a}
\end{equation}
For between-module movements, we use $q_{\curvearrowright}$ for the
rate of use of the index codebook with module code words used according
to the probability distribution $Q$. For within-module movements,
we use $p_{\circlearrowright}^{i}$ for the rate of use of the $i$-th
codebook with node and exit code words used according to the probability
distribution $\mathcal{P}^{i}$.

By minimizing the map equation over network partitions, we can resolve
how many modules we should use and which nodes should be in which
modules to best capture the dynamics on the network. See \url{http://www.mapequation.org}
for a dynamic visualization of the mechanics of the map equation.
Because the map equation only depends on the rates of node visits
and module transitions, it is universal to all flow for which the
rates of node visits and module transitions can be measured or calculated.
The code structure of the map equation can also be generalized to
make use of higher-order structures. In ref.~\cite{RosvallMultilevel2011},
we show how a multilevel code structure can reveal hierarchical organization
in networks, and in the next section, we show that we can capitalize
on overlapping structures by releasing the constraint that a node
can only belong to one module codebook.

\subsection{The map equation for overlapping modules}

The code structure of the map equation framework is flexible and can
be modified to uncover different structures of a network as long as
flow on the network can be unambiguously coded and decoded. As we
will show here, by releasing the constraint that a node can only belong
to one module codebook and allowing nodes to be \emph{information free
ports}, we can reveal overlapping modular organization in networks.
To see how, let us again study the air traffic between North America
and Europe in Fig.~\ref{fig:Illustration-of-hypothetical}(a). Suppose
that cities in North America and Europe belong to two different modules,
for simplicity identical in size and composition, and we are to assign
membership to Reykjavik between North America and 
Europe. For a hard partition, we would assign Reykjavik to the module
that most passengers travel to and from, and if the traffic flow were
the same, we could chose either module. But if the flow to and from
Reykjavik were dominated by American and European tourists visiting
Iceland for sightseeing before returning to their home continent,
both Americans and Europeans would consider Iceland as part of their
territory. We can accommodate for this view if we allow nodes to belong
to multiple module codebooks; depending on the origin of the flow,
we use different code words for the same node.

With the map equation for overlapping modules, we can measure the
description length of flow on the network with nodes assigned to multiple
modules. By minimizing the map equation for overlapping modules, we
can not only resolve into how many modules a network is organized
and which nodes belong to which modules, but also which nodes belong
to multiple modules and to what degree.

The pattern of flow, returning tourists to Iceland or in-transit businessmen
on intercontinental trips, determines whether we should assign Reykjavik
to North America, Europe, or both. Or, conversely, when we decide
whether Reykjavik should be assigned to North America, Europe, or
both, we reveal the pattern of boundary flow between modules, as Fig.~\ref{fig:Illustration-of-hypothetical} 
illustrates. In this hypothetical
example, assigning cities to two non-overlapping modules is always
better than assigning all cities to one module. But for a sufficiently
high proportion of returning flow, the overlapping modular solution
with Reykjavik in both modules as a free port provides the most efficient
partition to describe flow on the network.

The map equation for overlapping modules can take advantage of regularities
in the boundary flow between modules. To measure the length of an
overlapping modular description of flow on a network, we must decide
how the flow switches modules to calculate the node-visit rates from
different modules of multiply assigned nodes. In the Materials and
Methods section, we provide a detailed description of how a random
walker moves in an overlapping modular structure, but the rule is
simple: when a random walker arrives at a node assigned to multiple modules,
the walker remains in the same module if possible. Otherwise, the
random walker switches, with equal probability, to one of the modules
to which the node is assigned.

\begin{figure}
\includegraphics[width=1.0\columnwidth]{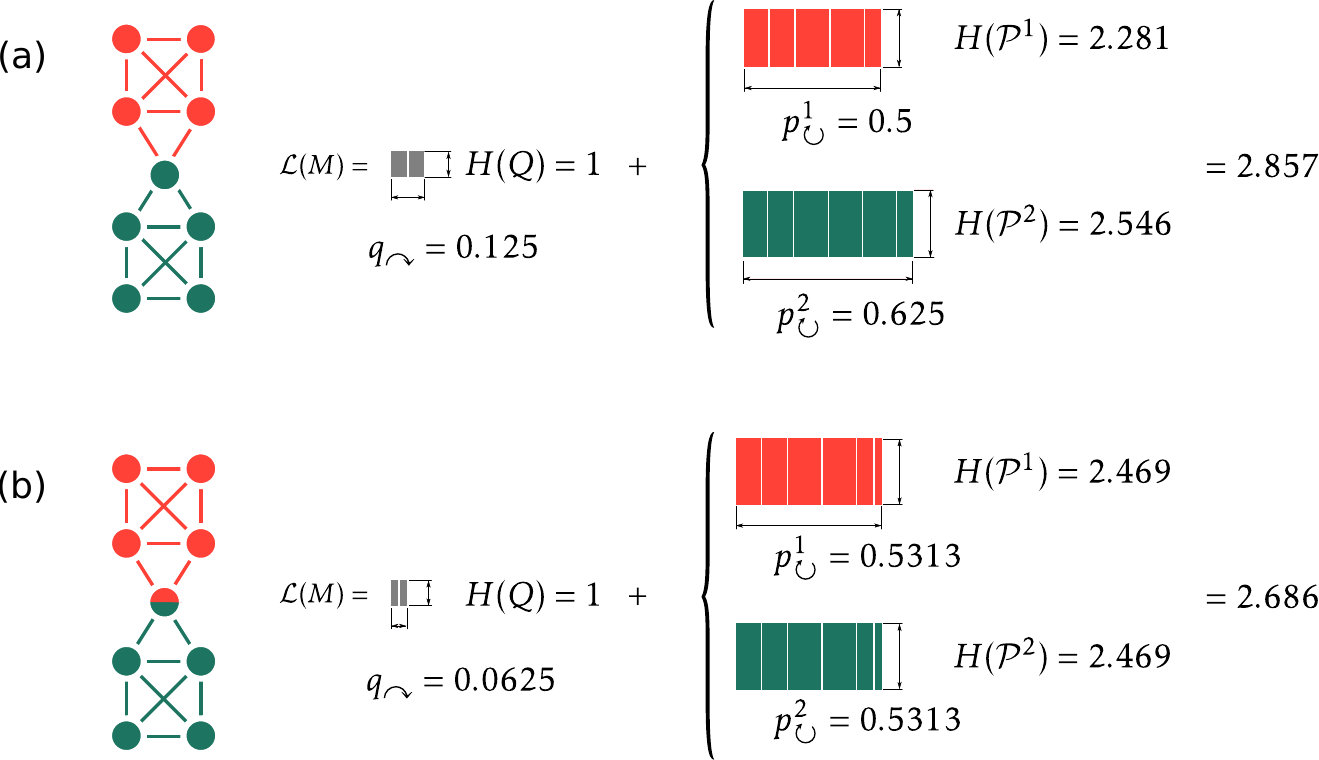}

\caption{\label{fig:Hard-versus-fuzzy}The code structure of the map equation
(a) without and (b) with overlapping modules. The color of a node
in the networks and of the corresponding block in the code structures
represents the module assignment, the width of a block represents
the node-visit rate, and the height of the blocks represents the average
codelength of code words in the codebooks. }
\end{figure}

Figure \ref{fig:Hard-versus-fuzzy} illustrates the code structure
of a hard and a fuzzy partition of an example network with the dynamics
derived from a random walker. For this network, the figure shows that
an overlapping modular description allows us to describe the path
of a random walker with fewer bits than we could do with a hard network
partition. With overlapping modules, we halve the use of the index
codebook, since the rate of module switching halves. Because we consequently
use the exit codes in the now identical module codebooks less often,
the description of movements within modules also becomes shorter,
even if the average code word length increases. Turning the reasoning
around again, given the overlapping modular organization, we have
learned that returning flow characterizes the boundary flow between
the modules. 

With the mathematical foundation in place, we need an algorithm that
can discover the best partition of the network. In particular, which
nodes should belong to multiple modules and to what degree? For this
optimization problem, we have developed a greedy search algorithm that
we call Fuzzy infomap and detail in the Materials and Methods section. Here we give a short
summary of Fuzzy infomap designed to provide good approximate
solutions for large networks. We start from Infomap's 
hard clustering of the network and then execute the two-step algorithm.
In the first step, we measure the change in the description length
when we assign boundary nodes, one by one, to multiple modules. This
calculation is fast, but aggregating the changes in the second step
is expensive and often requires recalculating all node-visit rates.
Therefore, we rank the individual multiple module assignments and,
in a greedy fashion, aggregate the individual best ones to minimize
the description length. 

\subsection{Overlapping modular organization in real-world networks}

\begin{figure*}
\includegraphics[height=0.9\textheight]{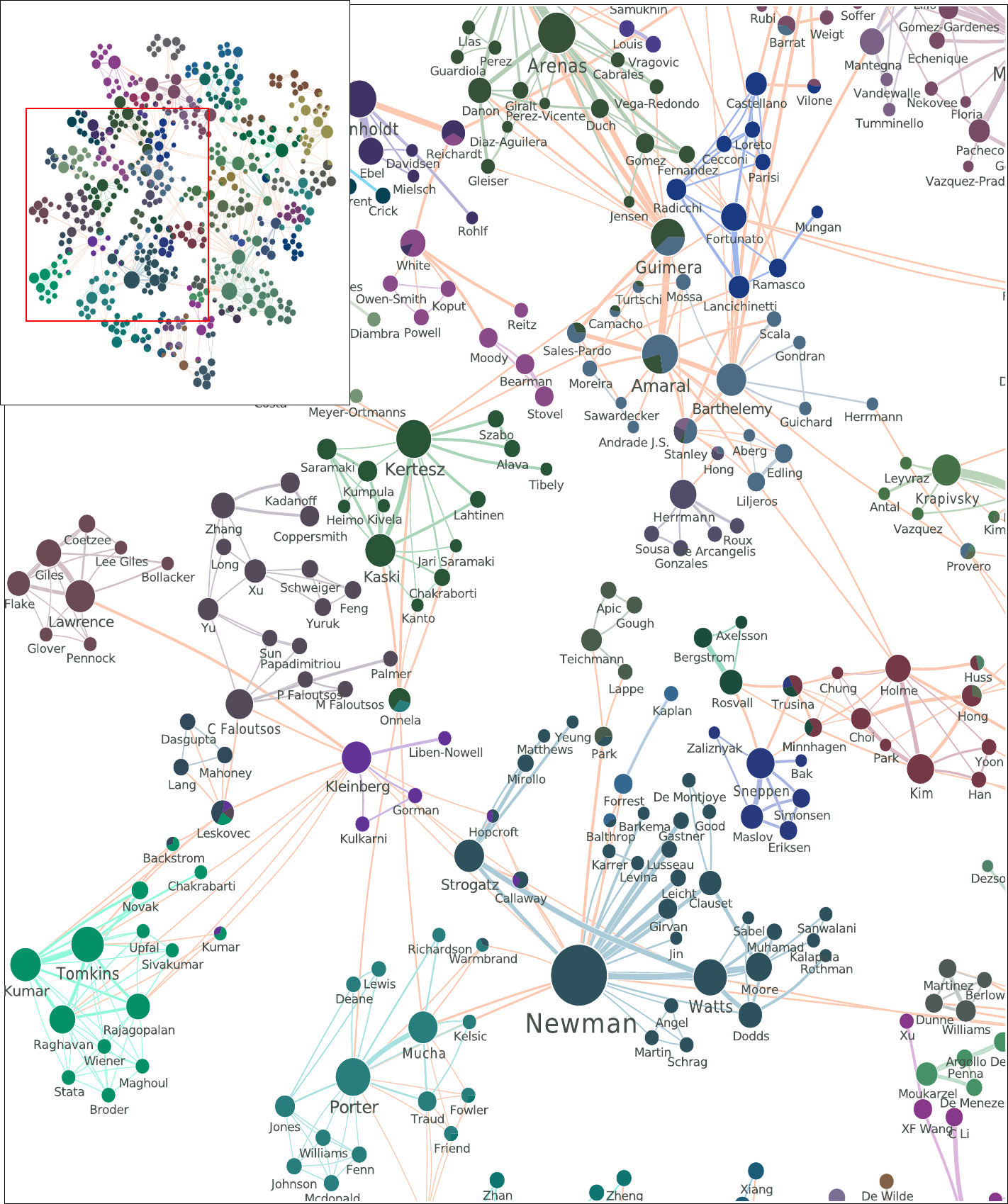}

\caption{\label{fig:Groups-are-better}Network scientists organized in overlapping
research groups. The colors of the nodes represent overlapping research
groups identified by the map equation and the pie charts represent
the fractional association with the different research groups.}
\end{figure*}

To illustrate our flow-based approach, we have clustered a number
of real-world networks. Figure \ref{fig:Groups-are-better} shows
researchers organized in overlapping research groups in network science.
The underlying co-authorship network is derived from the reference
lists in the three review articles \cite{albert2002,newmanSIAM,fortunato}.
In this weighted undirected network, we connect two researchers with
a weighted link if they have co-authored one or more research papers.
For every co-authored paper, we add to the total weight of the link
a weight inversely proportional to the number of authors on the paper.
Our premise is that two persons who have co-authored a paper have
exchanged information, information they can subsequently share with
other researchers and induce a flow of information on the network.
The map equation can capitalize on regularities in this flow, and
Fig.~\ref{fig:Groups-are-better} highlights one area of the co-authorship
network with several overlapping research groups. For example, assigning Jure
Leskovec to four research groups contributes to maximal compression
of a description of a random walker on the network. Based on this
co-authorship network, Leskovec is strongly associated with Dasgupta,
Mahoney, Lang, and Backstrom, but also with groups at
Cornell University, Carnegie Mellon University, Stanford University,
and Yahoo Research. The size of the modules and the fraction of returning
flow at the boundary nodes determine whether hard or fuzzy boundaries
between research groups lead to optimal compression of flow on the
network.

Table \ref{tab:Other-networks} shows the level of compression and
overlap of a number of real-world networks. The networks are sorted
from highest to lowest compression gain when allowing for overlaps.
We find the highest compression gain in the European roads network,
which is a sparse network with intersections as nodes and roads as
links. Many intersections at boundaries between modules are classified
in multiple modules, because intersections only connect a few roads
and the return rate of the random flow is relatively high. 

By contrast, compressing random flow in overlapping modules only gives
a marginal gain over hard clustering in the highly interconnected and
directed network of \emph{C.~Elegans}, where less than three percent
of the neurons are classified in multiple modules. Even if there is 
evidence that the neural network is modular, we most likely
underestimate the degree of overlap with a random walk model of flow.

In the middle of the table, the world air routes network shows a
relatively low compression gain, given the many cities classified in
multiple modules. For this network, the compression gain would be much
higher if, instead of random flow on the links, we were to describe
real passenger flow with a higher return rate.

\begin{table*}
\caption{\label{tab:Other-networks}The overlapping organization and the level
of compression of eight real-world networks. For each network with
$n$ nodes and $l$ links, we report the hard partition compression
$C$ with Infomap, the additional compression with Fuzzy infomap,
and the fraction of nodes that are assigned to multiple modules.}

\medskip{}

\begin{tabular*}{0.7\textwidth}{@{\extracolsep{\fill}} l r r D{.}{.}{-1} D{.}{.}{-1} D{.}{.}{-1} }
{ Network} & $n$ & $l$ & $C$ & \multicolumn{1}{c}{$\triangle C_{\textrm{fuzzy}}$} & \multicolumn{1}{c}{$N_{\textrm{fuzzy}}/N$}  \\
\hline
{European roads network\textsuperscript{\cite{eroadsnet}} } &{ 1018}&{ 1274} & 46.2\% & 10.4\% & 35.5\% \\
{Western states power grid\textsuperscript{\cite{watts1998collective}} } &{ 4941}&{ 6994} & 53.4\% & 8.84\% & 27.5\% \\
{Human diseases network\textsuperscript{\cite{goh2007human}} } &{ 516}&{ 1188} & 46.4\% & 2.87\% & 15.3\% \\
{Coauthorhip network\textsuperscript{\cite{coauthor}} } &{ 552}&{ 1317} & 48.9\% & 2.47\% & 14.6\% \\
{World air routes\textsuperscript{\cite{guimera2005worldwide}} } &{ 3618}&{ 14142} & 31.1\% & 1.24\% & 13.9\% \\
{U.S. political blogs\textsuperscript{\cite{adamic2005political}} } &{ 1222}&{ 16714} & 4.13\% & 0.35\% & 5.81\% \\
{Swedish political blogs\textsuperscript{\cite{blogs}} } &{ 855}&{ 10315} & 0.50\% & 0.18\% & 4.79\% \\
{Neural net. of \textit{C.\ Elegans}\textsuperscript{\cite{watts1998collective}} } &{ 297}&{ 2345} & 1.16\% & 0.13\% & 2.69\% \\
\hline 
\end{tabular*}
\end{table*}

\subsection{Comparing the map equation for overlapping modules with other methods}

Depending on the system being studied and the research question at hand, researchers
develop clustering algorithms for overlapping modules based on different principles.
For example, while some researchers take a statistical approach 
and see modules as non-random features of a network,
other researcher use a local definition and identify independent and
intersecting modules, or take a link perspective and assign all boundary nodes
to multiple modules.  Consequently, the final partitions are quite different,
and it is interesting to contrast our information theoretic and flow-based
approach, implemented in fuzzy infomap with these approaches, here represented
by OSLOM \cite{Lancichinetti2011}, Clique Percolation
\cite{palla2005uncovering}, and Link Communities \cite{Ahn:2010p167}.
 
OSLOM defines a module as the set of nodes that maximizes a
local statistical significance metric. In other words, OSLOM identifies
possibly overlapping modules that are unlikely to be found in a random network.
Clique percolation identifies clusters by sliding fully connected k-cliques to adjacent 
k-cliques that share k-1 vertices with each other.
A module is defined as the maximal set of nodes that can be visited in chained
iterations of this operation, and the overlaps consist of the
shared nodes between modules that do not support the slide operation across the boundary. 
Finally, the Link Communities approach creates highly overlapping modules by aggregating nodes that are
part of a link community. The link communities themselves are built using a
similarity measure between links, the primal actors of the method.

To compare the methods at different degrees of overlap,
we used a set of synthetic networks presented in ref.~\cite{lancichinetti2009benchmarks}.
In Table \ref{tab:all_the_methods}, we included six statistics for the
four methods applied to synthetic networks with 1000 nodes and three different degrees of overlap (see table caption for details).
The first group of partition numbers describe the number of detected 
modules, the number of nodes that are assigned to multiple modules,
and the total number of assignments. To interpret the results from a flow
perspective, we included the index, module, and total codelength
for describing a random walker on the network given the network partition.

\begin{table}

\begin{tabular*}{1\columnwidth}{@{\extracolsep{\fill}}l>{\centering}p{1cm}>{\centering}p{1cm}>{\centering}p{1cm}c>{\centering}p{0.8cm}>{\centering}p{0.8cm}>{\centering}p{0.8cm}}
 & \multicolumn{3}{c}{Partition numbers} &  & \multicolumn{3}{c}{Codelength (bits)}\tabularnewline
 & \texttt{\tiny modules} & \texttt{\tiny overlaps} & \texttt{\tiny assignments} &  & \texttt{\tiny index} & \texttt{\tiny module} & \texttt{\tiny total}\tabularnewline
\cline{2-4} \cline{6-8} 
\textbf{\scriptsize Low overlap} &  &  &  &  &  &  & \tabularnewline
{\scriptsize Fuzzy Infomap} & {\scriptsize 44} & {\scriptsize 105} & {\scriptsize 1228} &  & {\scriptsize 1.7} & {\scriptsize 5.9} & {\scriptsize 7.6}\tabularnewline
{\scriptsize OSLOM} & {\scriptsize 44} & {\scriptsize 89} & {\scriptsize 1089} &  & {\scriptsize 1.8} & {\scriptsize 5.8} & {\scriptsize 7.6}\tabularnewline
{\scriptsize Clique Percolation} & {\scriptsize 43} & {\scriptsize 104} & {\scriptsize 1108} &  & {\scriptsize 1.7} & {\scriptsize 6.0} & {\scriptsize 7.7}\tabularnewline
{\scriptsize Link Communities} & {\scriptsize 3415} & {\scriptsize 1000} & {\scriptsize 9215} &  & {\scriptsize 8.1} & {\scriptsize 3.5} & {\scriptsize 12}\tabularnewline
 &  &  &  &  &  &  & \tabularnewline
\textbf{\scriptsize Medium overlap} &  &  &  &  &  &  & \tabularnewline
{\scriptsize Fuzzy Infomap} & {\scriptsize 53} & {\scriptsize 303} & {\scriptsize 1830} &  & {\scriptsize 2.2} & {\scriptsize 6.0} & {\scriptsize 8.2}\tabularnewline
{\scriptsize OSLOM} & {\scriptsize 54} & {\scriptsize 276} & {\scriptsize 1277} &  & {\scriptsize 2.3} & {\scriptsize 5.9} & {\scriptsize 8.2}\tabularnewline
{\scriptsize Clique Percolation} & {\scriptsize 55} & {\scriptsize 268} & {\scriptsize 1283} &  & {\scriptsize 2.3} & {\scriptsize 6.1} & {\scriptsize 8.3}\tabularnewline
{\scriptsize Link Communities} & {\scriptsize 4457} & {\scriptsize 1000} & {\scriptsize 11628} &  & {\scriptsize 8.7} & {\scriptsize 3.5} & {\scriptsize 14}\tabularnewline
 &  &  &  &  &  &  & \tabularnewline
\textbf{\scriptsize High overlap} &  &  &  &  &  &  & \tabularnewline
{\scriptsize Fuzzy Infomap} & {\scriptsize 56} & {\scriptsize 398} & {\scriptsize 1676} &  & {\scriptsize 2.6} & {\scriptsize 6.1} & {\scriptsize 8.8}\tabularnewline
{\scriptsize OSLOM} & {\scriptsize 61} & {\scriptsize 462} & {\scriptsize 1465} &  & {\scriptsize 2.8} & {\scriptsize 6.0} & {\scriptsize 8.8}\tabularnewline
{\scriptsize Clique Percolation} & {\scriptsize 73} & {\scriptsize 388} & {\scriptsize 1429} &  & {\scriptsize 2.9} & {\scriptsize 6.1} & {\scriptsize 9.0}\tabularnewline
{\scriptsize Link Communities} & {\scriptsize 4298} & {\scriptsize 1000} & {\scriptsize 11063} &  & {\scriptsize 10} & {\scriptsize 3.7} & {\scriptsize 11}\tabularnewline
\end{tabular*}

\caption{\label{tab:all_the_methods} Comparing four different overlapping clustering methods.
We run Fuzzy infomap, OSLOM, and Link communities with their default settings
and use clique size four for the Clique percolation method.  All values are
averaged over ten instantiations of random undirected and unweighted networks
with 1000 nodes and predefined community structure, generated with three
different degrees of overlap \cite{lancichinetti2009benchmarks}: Low overlap
corresponds to 100, medium overlap corresponds to 300, and high overlap
corresponds to 500 nodes in multiple modules.  All other parameters were held
constant: The number of nodes that multiply-assigned nodes are assigned to was
set to two; each cluster consisted of on average 20 nodes with a minimum of
$10$ and a maximum of $50$ nodes; and the power law exponent was set to $-2$
for of the node degree distribution and $-1$ for the module size distribution.
Finally, the mixing parameter that controls the proportion of links within and
between modules was set to $0.1$.
}
\end{table}

Table \ref{tab:all_the_methods} shows that Fuzzy infomap and OSLOM
generate similar partitions for low and medium degrees of overlap,
but the trend when going to higher degrees of overlap indicates fundamental differences.
By assigning boundary nodes to more modules than OSLOM prefers,
Fuzzy infomap identifies modules with longer persistence times.
The shorter index codelength resulting from the fewer transitions 
compensates for the longer module codelength from the larger modules.
As a result, with the overlapping partitions generated by Fuzzy infomap,
random flow can be described with fewer bits. But the difference is 
small and shows up only in the second decimal place when up to half of all the 
nodes are assigned to multiple modules.

Clique percolation generates partitions with more modules but fewer assignments than both Fuzzy infomap and OSLOM.
From a flow perspective, smaller modules with less overlap give more module switches 
that cannot be compensated for by a shorter module codelength.
The strength of the Clique percolation method is the simple definition that 
allows for easy interpretation of the results.

Designed with links as the primal actors used to identify pervasive overlap in
networks, the results of Link Communities are quite different. For example,
independent of the degree of overlap of the synthetic networks, each node
belongs to on average ten modules.  From the perspective of a random flow
model, the persistence time is short in the many small modules, and the
information necessary to encode the many transitions is much larger than for
the other methods.  This result is expected, as Link Communities is tailored to
identify pervasive overlap in social networks in which people belong to several
modules and information flow is far from random.

Often  the performance is an important aspect to consider when choosing a
clustering method.  Therefore, we measured the time it took to cluster the
synthetic networks with the different clustering algorithms.  We stress that we
used presumably non-optimized research code made available online by its
developers and that the performance, of course, depends on the network.  Per 1000
node synthetic network used in our comparison, Fuzzy infomap used on average
1.7 seconds for a single iteration of module growth and 240 seconds for
multiple growths, OSLOM used 330 seconds, the Clique Percolation method 1.5
seconds, and link communities were identified in 2.4 seconds.

We conclude this comparison by stressing that the research question at hand
must be considered when choosing a clustering method.  Fuzzy infomap provides
fast results that, for a random flow model, are similar to results generated by
OSLOM and the Clique percolation method, at least for moderate degree of
overlap. On the other hand, for identifying pervasive overlap, researchers
should consider Link Communities or a generalized flow model with longer
persistence times in smaller, highly overlapping modules.

\section{\label{sec:Materials-and-methods}Materials and methods}

Here we detail the map equation for overlapping modules and describe
our greedy search algorithm. 

\subsection{The map equation for overlapping modules}

Below we explain in detail how we derive the transition rates of a
random walker between overlapping modules. We also derive the conditional
probabilities for nodes assigned to multiple modules. We then express
the map equation (Eq.~\ref{eq:a}) in terms of these rates, which
allows for fast updates in the search algorithm.

\subsubsection{Movements between nodes assigned to multiple modules\label{sub:How-the-random}}

To calculate the map equation for overlapping modules, we need the
visit rates $p_{\alpha_{i}}$ for all modules $i\in M_{\alpha}$ a
node $\alpha$ is assigned to and the inflow $q_{i\curvearrowleft}$
and the outflow $q_{i\curvearrowright}$ of all modules. We derive
these quantities from the weighted and directed links $W_{\alpha\beta}$,
which we normalize such that $w_{\alpha\beta}$ correspond to the
probability of the random walker moving to node $\beta$ once at node
$\alpha$:

\begin{equation}
w_{\alpha\beta}=\begin{cases}
0, & \mathrm{if\: there\: is\, no\: link\: from}\:\alpha\mathrm{\: to\:}\beta\\
\frac{W_{\alpha\beta}}{\underset{\beta}{\sum}W_{\alpha\beta}}, & \mathrm{otherwise}
\end{cases}.
\end{equation}

When necessary, we use random teleportation to guarantee a unique
steady state distribution \cite{google}. That is, for directed networks,
at rate $\tau$, or whenever the random walker arrives at a node with
no out-links, the random walker teleports to a random node in the
network. To simplify the notation, we set $w_{\alpha\beta}=1/n$ for
all nodes $\alpha$ without out-links to all $n$ nodes $\beta$ in
the network.

\begin{figure}
\includegraphics[width=0.4\columnwidth]{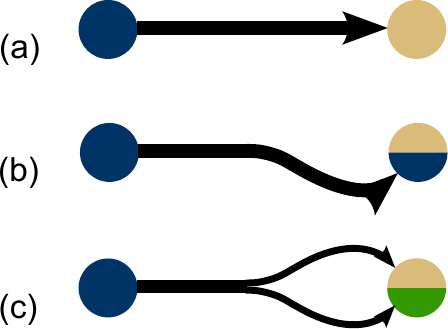}

\caption{\label{fig:Walker-behaviour-when}Movements between nodes possibly
assigned to multiple modules. (a) and (b) Assuming that the random walker
is in module $i$, she remains in module $i$ when moving to node
$\beta$ if node $\beta$ is assigned to module $i$. (b) But if node
$\beta$ is not assigned to module $i$, she switches with equal probability
to any of the modules node $\beta$ is assigned to. }
\end{figure}

The movements between multiply assigned nodes and overlapping modules
are straightforward. Whenever the random walker arrives at a node
that is assigned to multiple modules, she remains in the same module
if possible or switches to a random module if not possible. For example,
assuming that the random walker is in module $i$, she remains in
module $i$ when moving to node $\beta$ if node $\beta$ is assigned
to module $i$, $i\in M_{\beta}$. But if node $\beta$ is not assigned
to module $i$, $i\notin M_{\beta}$, she switches with equal probability
$1/\left|M_{\beta}\right|$ to any of the modules to which node $\beta$
is assigned(see Fig.~\ref{fig:Walker-behaviour-when}). If we define
the transition function

\begin{equation}
\delta_{\alpha_{i}\beta_{j}}=\begin{cases}
\begin{array}{l}
1,\\
\frac{1}{\left|M_{\beta}\right|},\\
0,
\end{array} & \begin{array}{l}
\mathrm{if}\: i=j\\
\mathrm{if\: i\neq j\:\mathrm{and\: i\notin M_{\beta}}}\\
\mathrm{if\: i\neq j\:\mathrm{and\: i\in M_{\beta}}}
\end{array}\end{cases},
\end{equation}
we can now define the visit rates by the equation system

\begin{equation}
p_{\alpha_{i}}=\underset{\beta}{\sum}\underset{j\in M_{\beta}}{\sum}p_{\beta_{j}}\delta_{\alpha_{i}\beta_{j}}\left[\left(1-\tau\right)w_{\beta\alpha}+\tau\frac{1}{n}\right].\label{eq:visit rate}
\end{equation}
We solve for the unknown visit rates with the fast iterative algorithm
BiCGStab\cite{BicGSTAB}. Since every node in module $i$ guides a fraction $\left(1-\tau\right)\sum_{\beta\notin i}w_{\alpha\beta}$
and teleports a fraction $\tau\frac{n-n_{i}}{n}$ of its conditional
probability $p_{\alpha_{i}}$ to nodes outside of module $i$, the
exit probability of module $i$ is

\begin{equation}
q_{i\curvearrowright}=\underset{\alpha\in i}{\sum}p_{\alpha_{i}}\left[\left(1-\tau\right)\underset{\beta\notin i}{\sum}w_{\alpha\beta}+\tau\frac{n-n_{i}}{n}\right],\label{eq:transition rate}
\end{equation}
where $n_{i}$ is the number of nodes assigned to module $i$.

\subsubsection{\label{sub:The-fuzzy-map}The expanded map equation for overlapping
modules}

To make explicit which terms must be updated in a given step of a
search algorithm, here we expand the entropies of the map equation
(Eq.~\ref{eq:a}) in terms of the visit and transition rates $p_{\alpha_{i}}$,
$q_{i\curvearrowleft}$, and $q_{i\curvearrowright}$. When teleportation
is included in the description length as above, the outflow of modules
balances the inflow, but here we derive for the general case when
$q_{i\curvearrowleft}\neq q_{i\curvearrowright}$.

We use the per-step probabilities of entering the modules $q_{i\curvearrowleft}$
to calculate the average code word length of the index code words weighted
by their rates of use, which is given by the entropy for the index
codebook

\begin{equation}
H\left(\mathcal{Q}\right)=-\sum_{i=1}^{m}\frac{q_{i\curvearrowleft}}{\sum_{j=1}^{m}q_{j\curvearrowleft}}\log_2\frac{q_{i\curvearrowleft}}{\sum_{j=1}^{m}q_{j\curvearrowleft}},\label{eq:index-codebook-entropy}
\end{equation}
where the sum runs over the $m$ modules of the modular partition.
The contribution to the average description length from the index
codebook is the entropy $H\left(\mathcal{Q}\right)$ weighted by its
rate of use $q_{\curvearrowleft}$, 
\begin{equation}
q_{\curvearrowleft}=\sum_{j=1}^{m}q_{j\curvearrowleft}.\label{eq:sum-of-entries}
\end{equation}
 Substituting Eq.~\ref{eq:sum-of-entries} into Eq.~\ref{eq:index-codebook-entropy},
we can express the contribution to the per-step average description
length from the index codebook as

\begin{eqnarray}
q_{\curvearrowleft}H\left(\mathcal{Q}\right) & = & -q_{\curvearrowleft}\left[\sum_{i=1}^{m}\frac{q_{i\curvearrowleft}}{q_{\curvearrowleft}}\log_2\frac{q_{i\curvearrowleft}}{q_{\curvearrowleft}}\right]\nonumber \\
 & = & -\sum_{i=1}^{m}q_{i\curvearrowleft}\left[\log_2 q_{i\curvearrowleft}-\log_2 q_{\curvearrowleft}\right]\nonumber \\
 & = & q_{\curvearrowleft}\log_2 q_{\curvearrowleft}-\sum_{i=1}^{m}q_{i\curvearrowleft}\log_2 q_{i\curvearrowleft}.\label{eq:index_codebook_term}
\end{eqnarray}

We use the per-step probabilities of exiting the modules $q_{i\curvearrowright}$
and the visit rates $p_{\alpha_{i}}$ to calculate the entropy of
each module codebook: 
\begin{multline}
H\left(\mathcal{P}^{i}\right)=-\frac{q_{i\curvearrowright}}{q_{i\curvearrowright}+\sum_{\beta\in i}p_{\beta i}}\log_2\frac{q_{i\curvearrowright}}{q_{i\curvearrowright}+\sum_{\beta\in i}p_{\beta i}}-\\
-\sum_{\alpha\in i}\frac{p_{\alpha_{i}}}{q_{i\curvearrowright}+\sum_{\beta\in i}p_{\beta i}}\log_2\frac{p_{\alpha_{i}}}{q_{i\curvearrowright}+\sum_{\beta\in i}p_{\beta i}}\\
=-\frac{1}{p_{\circlearrowright i}}\left[q_{i\curvearrowright}\log_2 q_{i\curvearrowright}+\sum_{\alpha\in i}p_{\alpha_{i}}\log_2 p_{\alpha_{i}}-p_{\circlearrowright}^{i}\log_2 p_{\circlearrowright}^{i}\right],
\end{multline}
with $p_{\circlearrowright}^{i}$ for the rate of use of the $i$-th
module codebook,

\begin{equation}
p_{\circlearrowright}^{i}=q_{i\curvearrowright}+\sum_{\beta\in i}p_{\beta_{i}}.
\end{equation}

Finally, summing over all module codebooks, the description length
given by the overlapping module partition $\mathsf{M}$ is 
\begin{eqnarray}
L\left(\mathsf{M}\right) & = & q_{\curvearrowleft}\log_2 q_{\curvearrowleft}\nonumber \\
 &  & -\sum_{i=1}^{m}q_{i\curvearrowleft}\log_2 q_{i\curvearrowleft}-\sum_{i=1}^{m}q_{i\curvearrowright}\log_2 q_{i\curvearrowright}\\
 &  & -\sum_{i=1}^{m}\underset{\alpha\in i}{\sum}p_{\alpha_{i}}\log_2 p_{\alpha_{i}}+\sum_{i=1}^{m}p_{\circlearrowright}^{i}\log_2 p_{\circlearrowright}^{i}.\nonumber 
\end{eqnarray}
The only visible difference between this expression and the map equation
for non-overlapping modules is the sum over conditional probabilities
for nodes assigned to multiple modules, which is no longer independent
of the overlapping module partition $\mathsf{M}$. But since the transition
rates depend on the conditional probabilities (see Eq.~\ref{eq:transition rate}),
all terms depend on the overlapping configuration.

\subsection{\label{sub:The-optimization-method}The greedy search algorithm for
overlapping modules}

\begin{figure*}

\includegraphics[width=1\textwidth]{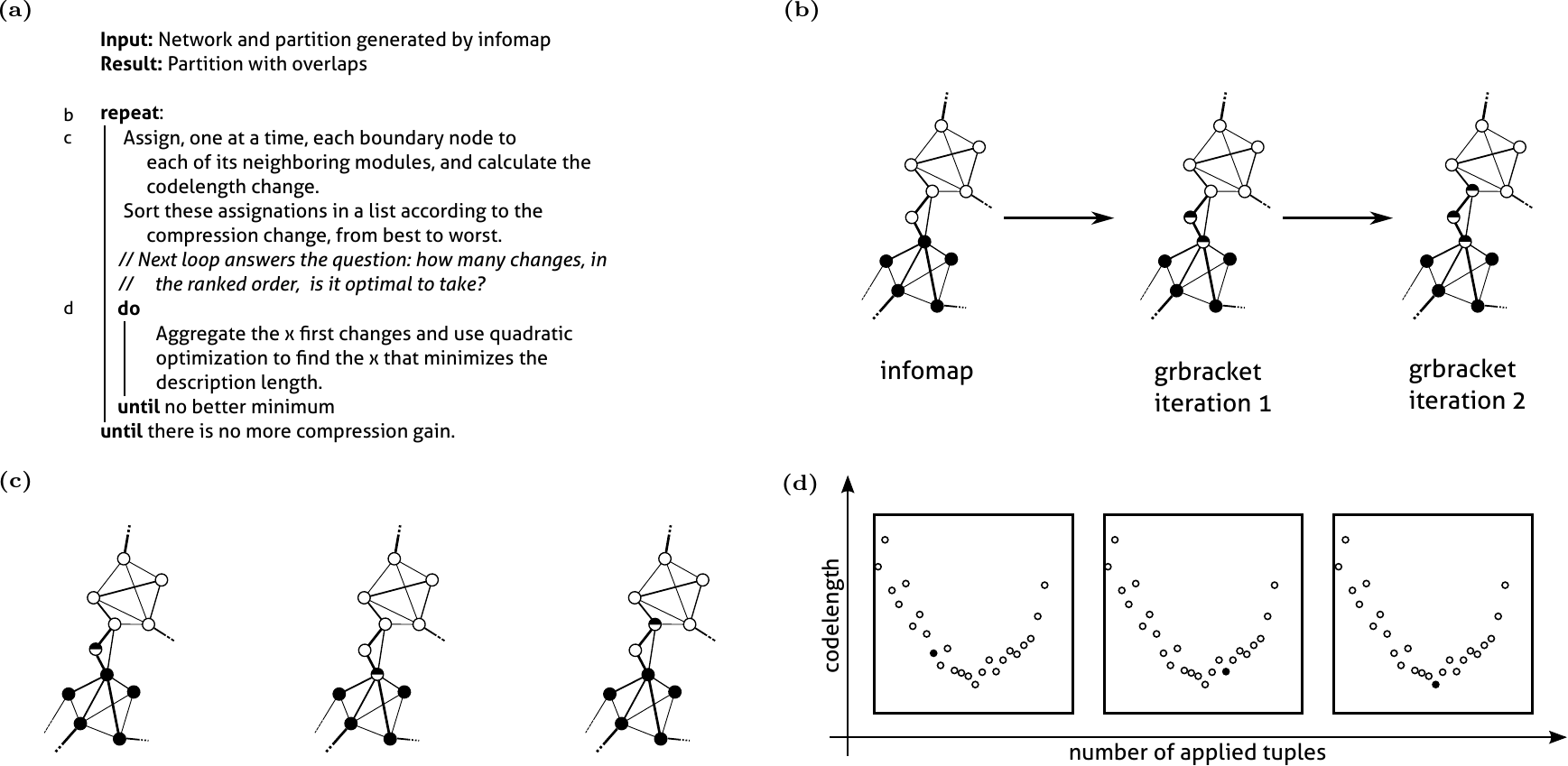}

\caption{\label{fig:General-scheme-of}General scheme of the two-step greedy
search algorithm for overlapping modules. (a) Pseudocode with first
step (c) and second step (d) of the algorithm that can be iterated
as shown in (b). Starting from a hard partition generated by Infomap
\cite{rosvall2009map}, each iteration successively increases the
overlap between modules to minimize the map equation for overlapping
modules. In the first step (c), one by one, each boundary node is
assigned to adjacent modules. In the second step (d), we first sort
the local changes from best to worst and then iteratively apply quadratic
fitting to find the number of best local changes that minimizes the
map equation.}
\end{figure*}

To detect the overlapping modular organization of a network, ultimately
we want to find the global minimum of the map equation over all possible
overlapping modular configurations of the network, but only with an
exhaustive enumeration of all possible solutions can we guarantee
the optimal solution. This procedure is, of course, impractical for
all but the smallest networks. However, we can construct an algorithm
that finds a good approximation. Figure \ref{fig:General-scheme-of}
explains the concept of our algorithm, which builds on an iterative
two-step procedure.

In the first step, we individually assess which nodes are most likely
to be assigned to multiple modules. Starting from a hard partition
generated by Infomap\cite{rosvall2009map} in the first iteration,
we go through all nodes at the boundary between modules and assign
each boundary node to adjacent modules. That is, one node and one
adjacent module at a time, we assign the node to the extra module,
measure the map equation change, and then return to the previous configuration
(see Fig.~\ref{fig:General-scheme-of}(c)). Because the multiply
assigned nodes only connect to singly assigned nodes in the first
iteration, the conditional probabilities and the change in the map
equation can be updated quickly without a full recalculation of the
visit rates. This first step produces 3-tuples of local changes of
the form \emph{(node, extra-module, map-equation-change)}.

In the second step, we combine a fraction of all local changes generated
in the first step into a global solution. Every time two or more multiply
assigned nodes are connected, we need to solve a linear system to
calculate the conditional probabilities. When a majority of nodes
are assigned to multiple modules, this can take as long as calculating
the steady-state distribution of random walkers in the first place.
For good performance, we therefore try to test as few combinations
of local changes as possible. After testing several different approaches,
we have opted for a heuristic method in which we first sort the tuples
from best to worst in terms of map equation change and then determine
the number of best tuples that minimizes the map equation. The method
works well, because good local changes often are good globally. 

As a side remark, the map equation for link community \cite{kim2011} allows for straightforward
and fast calculation of all conditional probabilities and transition
rates, since each link belongs to only one module. But this constraint
enforces module switches between boundary nodes that belong to the
same module, because all boundary nodes belong to multiple modules
in the link community approach.

Figure \ref{fig:General-scheme-of}(d) shows the value of the map
equation as a function of the number of aggregated tuples ordered
from best to worst. Combinations of tuples that individually generate
longer description lengths can generate a shorter description length
if they are applied together. This fact, together with the greedy
order in which we aggregate the tuples, generates noise in the curve.
To quickly approach the global minimum, we must overcome bad local
minima caused by the noise and evaluate as few aggregations as possible.
Therefore, we iteratively fit a quadratic polynomial to the curve
by selecting new points at the minimum of the polynomial. A quadratic
polynomial only requires three points to be fully specified, but in
order to deal with the noise, we use a moving local least squares fit. 
In practice, we evaluate around ten points for each quadratic fit and repeat
this procedure a few times to obtain a good solution.

Step 1 and step 2 can now be repeated, each time starting from the obtained
solution with overlapping modules from the previous iteration.  Figure
\ref{fig:General-scheme-of}(b) illustrates that by repeating the two steps, we
sometimes can extend the overlap between modules, but this comes at a cost.
After the first iteration of the algorithm, step 1 also can involve solving a
linear system to calculate the conditional probabilities. Thus, the first step
is no longer guaranteed to be as fast as in the first iteration. Still, for
medium-sized networks, multiple iterations are feasible. For example, for the
networks presented in Table \ref{tab:Other-networks}, the first iteration took
a few seconds and multiple iterations until the point of no further
improvements took less than two minutes on a normal laptop. We have made the
code available here: \url{https://sites.google.com/site/alcidesve82/}.

\section{Conclusions}

In this paper, we have introduced the map equation for overlapping
modules. When we allow nodes to belong to multiple module codebooks
and minimize the map equation over possibly overlapping network partitions,
we can determine which nodes belong to multiple modules and to what
degree. Compared to hard partitions detected by the map equation,
we have further compressed descriptions of a random walker on all
tested real-world networks, and therefore revealed more regularities
in the flow on the networks. We find the highest overlapping modular
organization in sparse infrastructure networks, but this result depends
on our random-walk model of flow. Since the mathematical framework
is not limited to random flow, it would be interesting to compare
our results with results derived from empirical flow.
\begin{acknowledgments}
We are grateful to Klas Markström and Daniel Andrén for several good
algorithmic suggestions. MR was supported by the Swedish Research
Council grant 2009-5344.
\end{acknowledgments}

\end{document}